\documentclass[aps,amsfonts,pra,showpacs,nobibnotes,nofootinbib,tightenlines,floatfix]{revtex4} 
\usepackage{graphicx}
\begin{document}

\newcommand{\beq}{\begin{equation}}
\newcommand{\eeq}{\end{equation}}
\newcommand{\barr}{\begin{eqnarray}}
\newcommand{\earr}{\end{eqnarray}}
\def\figwidth{7.5cm}

\newcommand{\Mfunction}[1]{#1}
\def\cD{{\cal D}}
\def\cF{{\cal F}}
\def\cE{{\cal E}}
\def\bfn{{\bf n}}

\def\Ord#1{{\cal O}\left( #1\right)}
\renewcommand{\Im}{{\rm Im}}


\title{The Casimir Energy for a Hyperboloid Facing a Plate in the
Optical Approximation}
\author{O.~Schr\"oder} \email{oliver.schroder@plymouth.ac.uk}
\affiliation{School of Mathematics and
  Statistics, University of Plymouth, Plymouth PL4 8AA, UK\\} 
\affiliation{Center for Theoretical Physics, \\ Laboratory for
Nuclear Sciences and Department of Physics\\
Massachusetts Institute of Technology \\ Cambridge, MA 02139, USA}
\author{A.~Scardicchio}\email{scardicc@mit.edu}

\author{R.~L.~Jaffe}\email{jaffe@mit.edu}
\affiliation{Center for Theoretical Physics, \\ Laboratory for
Nuclear Sciences and Department of Physics\\
Massachusetts Institute of Technology \\ Cambridge, MA 02139, USA}
\begin{abstract}
\noindent We study the Casimir energy of a massless scalar field
that obeys Dirichlet boundary conditions on a hyperboloid facing a
plate. We use the optical approximation including the first six
reflections and compare the results with the predictions of the
proximity force approximation and the semi-classical method.  We
also consider finite size effects by contrasting the infinite with
a finite plate.  We find sizable and qualitative differences
between the new optical method and the more traditional
approaches.
\end{abstract}
\pacs{03.65Sq, 03.70+k, 42.25Gy\\ [2pt] MIT-CTP-3578}
\vspace*{-\bigskipamount} \preprint{MIT-CTP-3578} \maketitle

\section{Introduction}
\setcounter{equation}{0} \setcounter{figure}{0}

The last ten years have seen a revolution in experimental
techniques used to measure the Casimir effect \cite{Casimir,
expt1, MPnew}.  These new techniques open the door to measurements at a
precision where interesting geometrical dependence can be expected
\cite{MPnew}.  However no exact calculations are available for
geometries other than parallel plates. Traditionally the only tool
for estimating Casimir energies for other geometries has been the
proximity force approximation (PFA), which treats all geometries
as superpositions of infinitesimal parallel plates
\cite{Derjagin}, a crude approximation.

It is therefore interesting to develop approximations that might
provide an accurate estimate of the Casimir energy for other,
experimentally relevant geometries.  Recently we have developed a
new, approximate treatment of Casimir effects for sufficiently
smooth but otherwise arbitrary geometries based on geometric
optics \cite{pap1,opt1}.  We have tested the optical method by
comparing with a precise numerical calculation \cite{Gies03} for
the case of a sphere facing a plate.  The optical approximation
agrees much better with the numerical results than does the PFA.
So far, it has not been possible to provide a useful estimate of
the corrections to the optical approximation, which involve
diffractive contributions.

Some years ago Schaden and Spruch proposed a``semi-classical"
approximation based on Gutzwiller's approximation for the density
of states \cite{SandS,Gutz}. This approximation treats high
frequency effects correctly and is exact for planar surfaces.  It
also captures important effects of curvature, which are clearly
omitted in the PFA.  However, as we will see, it does not
seem to capture other important aspects of the geometry.  For
example, it is sensitive only to the curvature of the boundaries
at their points of closest approach, whereas the general form of
the problem suggests much more complex dependence on the geometry.

In this paper we apply the optical approach to the study of the
experimentally relevant example of a hyperboloid facing a plate.
We consider a scalar field and impose Dirichlet boundary
conditions.  We are interested in the force between the
hyperboloid and the plate, or equivalently, the interaction
energy, from which divergent self-energies that do not contribute
to the force have been subtracted. This problem has no closed,
analytic solution, nor has it been studied with the numerical
methods of Gies {\it et al.\/}  \cite{Gies03} but it
seems like a good candidate
for future experimental studies.  We find that the optical
estimate of the Casimir energy differs significantly from the
other approximations especially when the opening angle of the
hyperboloid is small.

The optical method can be applied also when the bounding surfaces
are finite, so in order to assess the effects of finiteness we
study the configuration of a hyperboloid opposite to a finite
plate. We find and explain differences between the optical
approach on the  one hand and the PFA and the semi-classical
approach of Ref.~\cite{SandS} on the other.   This application
illustrates the shortcomings of the semi-classical approximation
and the more subtle difference between the PFA and the optical
approximation.  In particular it helps clarify in which sense the
optical approximation is a \emph{uniform} semi-classical
approximation.

The broad interest in  Casimir physics and the application of new
experimental methods will certainly allow tests which can distinguish
among these different approaches, and guide theory toward a correct
treatment of the dependence of Casimir effects on geometry.

\section{The Optical Approach to Casimir Energies}

We want to calculate the Casimir energy of a quantized scalar field
obeying boundary conditions on the border $\partial\cD$ of the domain
$\cD$ limited by impenetrable bodies.  This is an idealization of a
physical interaction that prevents the field from entering the bodies.
In the case of physical interest the electromagnetic field interacts
with the electrons in metallic bodies.  The interactions can be
idealized by conducting boundary conditions for momenta $k\ll \Lambda$
where $\Lambda$ is a cutoff.  For the case of the metal and
electromagnetic field, the cutoff is of order of the plasma
frequency of the material.  The Casimir energy depends on
 $\Lambda$, and would diverge if $\Lambda$ were taken
to infinity.  So the cutoff cannot be removed in the fashion
familiar from renormalizable quantum field theories.  This is not
a problem for us, however.  First, the would-be divergences are
associated with the self-energies of the bodies and do not
contribute to \textit{forces} (or \textit{interaction energies})
between \textit{rigid} bodies \cite{Graham:2003ib}, which are what
concerns us here.  Second, finite cutoff dependence can be ignored
when the minimum distance between the two bodies, $h$, is much
larger than the inverse cutoff, i.e. $h \gg 1/\Lambda$
\cite{pap1,opt1}.

The Casimir energy for a massless scalar field $\phi$ living
inside the domain $\cD\subset \mathbb{R}^3$ with Dirichlet
boundary conditions on the surface $\partial\cD$ can be written as
\cite{Graham:2002xq}
\begin{eqnarray}
    \label{eq:casimiren}
    \cE_{\cD}[\phi]=  \frac{1}{2}\hbar\ \int_0^\infty dk \int_{\cD}
    d^{3} x \ \omega(k)\rho(x,k),
\end{eqnarray}
where in the case of massless fields $\omega(k)=ck$, and the
(local) density of states $\rho(x,k)$ is related to the propagator
$G(x',x,k)$ of the Helmholtz equation by
\begin{equation}
    \label{eq:rhoG}
    \rho(x,k)=\frac{2k}{\pi}\ \Im \ G(x,x,k),
\end{equation}
and the standard density of states is $\rho(k)=\int d^3x\ \rho(x,k)$.
The equation satisfied by $G$ is
\begin{eqnarray}
\label{eq:Helmholtz}
(-\Delta'-k^2)G(x',x,k)&=&\delta^3(x'-x)\ \mbox{if}\
x,x'\in\cD\nonumber\\
G(x',x,k)&=&0\ \mbox{if
} x'\mbox{ or } x\in\partial\cD.
\end{eqnarray}

The essence of the optical approximation is to replace the
Helmholtz propagator, eq.\ (\ref{eq:Helmholtz}), with an
approximation taken from wave optics \cite{pap1,opt1} which
assumes that the path integral representation for $G(x,x',k)$ is
saturated by its stationary points, i.e.\ straight line paths
making specular reflections (accompanied by a phase change) at the
boundaries.  In this way the intractable sum over modes is
replaced by a tractable, but approximate sum over paths,
\begin{equation}
        {\cal E}_{\rm optical}=-\frac{\hbar c}{2\pi^{2}}\sum_{\bf
        n}(-1)^{n} M_{n}\int_{{\cal D}_{\bf
        n}}d^{3}x\frac{\sqrt{\Delta_{\bf n}(x)}}{\ell_{\bf
        n}^{3}(x)}\, .
        \label{eq2}
\end{equation}
Here the sum runs over the optical paths indexed by $\bfn$ (which
is an index taking care of both the number of reflections $n$ and
the sequence of bodies on which the reflections occur),
$\ell_\bfn(x)$ is the length of the closed path starting and
ending at $x$ and $\cD_{\bfn}$ is its domain of existence (which
can be smaller than $\cD$), $M_n$ is the multiplicity of the path
$\bfn$ ($M_n=1$ for paths with an odd number of reflections,
$M_n=2$ for paths with an even number of reflections) and
$\Delta_\bfn(x)$ is shorthand for the enlargement factor
$\Delta_\bfn(x,x)$:
\begin{equation}
    \label{eq:deltar}
    \Delta_\bfn(x',x)=\frac{d\Omega_{x}}{dA_{x'}}
\end{equation}
is the ratio between the angular opening of an arbitrarily narrow
pencil of rays following the optical path $\bfn$ starting at the
initial point $x$ and the area spanned at the final point
$x'$\cite{KandK}.

The origins of the optical approximation and further discussion of the
derivation and significance of quantities like the enlargement factor
can be found in Ref.~\cite{opt1}.  All the quantities that appear in
eq.~(\ref{eq2}) can be calculated numerically for any number of
reflections.  The details of the algorithm are sketched in the
appendix.

\section{Hyperboloid facing a plate}
\subsection{Parametrization}

\begin{figure}
\includegraphics[width=4cm]{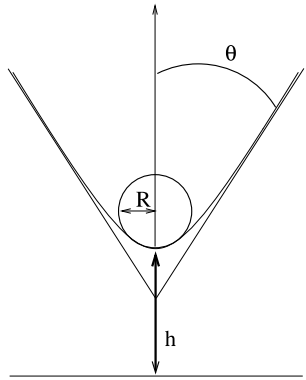}
\caption{\label{fig:v}\sl Configuration of hyperboloid and plate,
  illustrating the meaning of variables}
\end{figure}

A general parametrization of a cylindrically symmetric hyperboloid
centered on the $z$ axis a distance $h$ above the $z=0$ plate is given
by
\begin{equation}
z(\rho)=h-b+b\sqrt{1+\rho^2/a^2},
\end{equation}
where the parameters $b$ and $a$ measure the opening angle
$\theta$ and the radius of curvature $R$ according to
\begin{eqnarray}
R&=&a^2/b,\nonumber\\
\cot \theta&=&b/a.
\end{eqnarray}
The configuration is shown in Fig.\ \ref{fig:v}. We choose units
such that $\hbar=c=1$, and study the Casimir effects as functions
of the variables $h$, $R$ and $\theta$.  The limit $\theta\to 0$
at fixed $h$ and $R$ gives a paraboloid, $z=h+\rho^2/2R$, while
$\theta\to\pi/2$ with finite $a$ is the planar limit.

\subsection{Proximity force and semi-classical
  approximations}

The proximity force approximation is a first approximation to the
problem for arbitrary surfaces and is believed to give \emph{the most
divergent term} correctly in the limit of small distances, even though
no rigorous proof for this exists.  The PFA estimate is obtained by
integrating the parallel plate result,
\begin{equation}
\frac{d\cE}{dS}=-\frac{\pi^2}{1440z^3}
\end{equation}
over one of the surfaces with $z$ replaced by the distance to the
second surface measured normal to the first. In general, the PFA is
ambiguous, since there are two choices for the integration surface and
they yield, generically, different results.
We choose to integrate over the planar surface, and
obtain for the hyperboloid,
\begin{equation}
\label{eq:definef} \cE_{\rm PFA}=-\int dS\frac{\pi^2}{1440
z(\rho)^3}=-\frac{\pi^3R}{1440\
h^2}\left(1+\frac{h}{R}\tan^2\theta\right)\equiv
-\frac{\pi^3R }{1440\ h^2}f_{\rm PFA}(h/R,\theta),
\end{equation}
and we use the function $f(h/R, \theta)$ to present our results in
general, so
\begin{equation}
\label{eq:fpa}
f_{\rm PFA}(h/R,\theta)=1+\frac{h}{R}\tan^2\theta \, .
\end{equation}
The conjecture that the PFA captures correctly the most divergent
contribution in the limit $h\to 0$ translates to the conjecture
that $f(0,\theta)=1$ is exact.  We will see that both the optical
and semi-classical approximations reproduce this relation.

A semi-classical approximation for the Casimir energy has been
developed by Schaden and Spruch \cite{SandS} following Gutzwiller's
methods \cite{Gutz}.  Like the optical approach this method
identifies \textit{closed classical} paths in the propagator $G$
and expands the functional integral about them. Unlike the optical
approach, the trace of $G$ is calculated then by stationary phase
leaving only \textit{periodic} paths. In contrast, the optical
approximation uses all \textit{closed} and not necessarily
periodic paths. This should be \emph{almost} equivalent when only
almost periodic paths contribute; however there are situations in
which this is not true (see section \ref{sec:finite}) and
situations in which periodic paths do not exist, while closed
paths do (inside a wedge, for example). The approaches are
compared further in Ref.~\cite{opt1}. The resulting semi-classical
expression for the Casimir energy depends only on the local
properties of the surface in the neighborhood of the points of
reflection.  For the hyperboloid, there is only one periodic path,
the one that originates at the tip of the hyperboloid.  In this
case the approach of Schaden and Spruch gives
\barr
\label{eq:fss}
f_{\rm SS}(h/R,\theta)&=&\frac{90}{\pi^4} \frac{h}{R}
\sum_{n=1}^\infty\frac{1}{n^2\sinh^2\left(n\,\log \left(\sqrt{1 +
\frac{h}{R}} +
         \sqrt{\frac{h}{R}}\right)\right)\,
  }\nonumber \\
&=&1-\frac{30}{\pi^4}\sum_{n=1}\left(\frac{1}{n^2}-\frac{1}{n^4}\right)
\frac{h}{R}+\Ord{(h/R)^{3/2}}\nonumber\\
  &=&1-\frac{15-\pi^2}{3\pi^2}\frac{h}{R}+\Ord{(h/R)^{3/2}} \, .
\earr
Notice that this result depends on $R$ but not on the angle
$\theta$. In the semi-classical approximation the hyperboloid
gives the same result as a sphere of radius of curvature $R$,
irrespectively of the opening angle $\theta$.  The next term in
the power series is actually $\Ord{(h/R)^{1+\alpha}}$ with
$0<\alpha<1$.  Numerically, we found $\alpha=1/2$.  Notice that the
term proportional to $h/R$ has the opposite sign from our computations
and from the PFA prediction.

{\it A priori\/} we have no reason to dismiss either the PFA or
the semi-classical approximation.  Neither the $h/R$ nor the
$\theta$ dependence of $f(h/R,\theta)$ is constrained by any
general requirement.  The only test we can foresee is either
comparison with experiment or with a numerical computation after
the manner of Ref.~\cite{Gies03}. We believe that the optical
approximation captures more of the relevant physics than either
the PFA or semi-classical approximation.  The PFA ignores the
curvature of the surfaces entirely and the semi-classical
approximation ignores the geometry except in the neighborhood of
the periodic paths.  We have already seen \cite{opt1} in the case
of the sphere and the plate that the optical approximation gives a
prediction for the coefficient of the linear term $h/R$ (where $R$
is the radius of the sphere) different from either the PFA or
semi-classical approximations. 
The optical approximation differs significantly from the other
approaches also in the case of a hyperboloid (the difference being
more evident the smaller is the angular opening $\theta$ of the
hyperboloid), so experiments or numerical computation will again
have to provide discrimination among the approximations.

\subsection{Optical approach data}

\begin{figure}
\includegraphics[width=11cm]{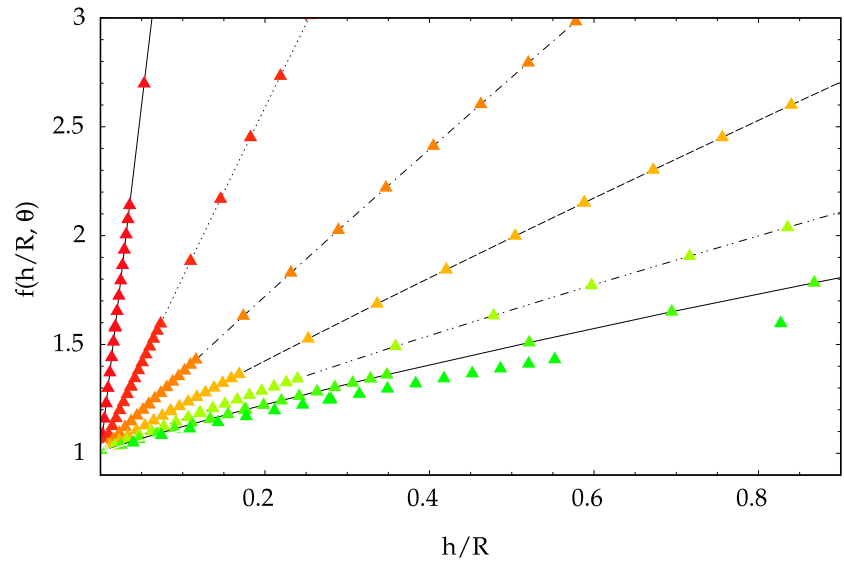}
\caption{\label{fig:f}\sl (Color online) Optical approximation for the
  function $f(h/R,\theta)$ versus\ $h/R$ for various openings of the
hyperboloid $\theta$.  From bottom up, azure to red, $\theta=20$
to $80$ degrees in steps of $10$ degrees.  The function
$A(\theta)$ is given by the slope at the origin.}
\end{figure}

\begin{figure}
\includegraphics[width=10cm]{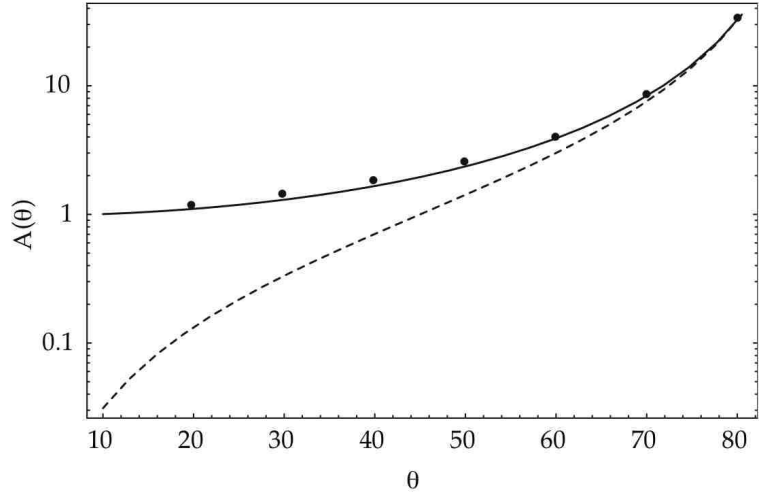}
\caption{\label{fig:a}\sl Optical approximation and PFA for
$A(\theta)$.  The continuous line is the function
$1/\cos^2\theta$, the dashed line is the PFA prediction
$\tan^2\theta$. $\theta=90^o$ is the parallel plates limit and the
optical approximation agrees in its functional form with the PFA
for large opening angles. The semi-classical prediction is
$A=-0.17$ which cannot be displayed on a logarithmic scale.}
\end{figure}

We have computed the energy in the optical approximation up to 6
reflections with a numerical algorithm (see Appendix for details)
for 7 different values of  $\theta$  (from 20 to 80  degrees) and
approximately 50 values of $h/R$ for each $\theta$. The data are
presented in Fig. \ref{fig:f}.
For small $h/R$ we can expand $f$:
\begin{equation}
\label{eq:defA}
f(h/R,\theta)=1+A(\theta)h/R +  \ldots \quad .
\end{equation}
This defines the function $A(\theta)$, which, for dimensional reasons,
can only depend on $\theta$.  The PFA gives only a term linear in
$h/R$, $A_{{\rm PFA}}(\theta)=\tan^2\theta$.  From Fig.~\ref{fig:f} it
is apparent that the optical approximation also is nearly linear in
$h/R$ over the range of $h/R$ and $\theta$ shown.  We have extracted
this function from our data and plot it in Fig.~\ref{fig:a}.

As can be seen from the figure, the results of the optical
approximation are well described by the function
\beq
\label{eq:opticalA}
A(\theta)=\frac{1}{\cos^2\theta} \, .
\eeq
So far we have no explanation for this functional form.  However,
it is so simple and fits the data so accurately that it has
probably a more profound meaning. To gain some further insight into
this subject consider the limit $\theta\to 0$ at fixed finite $R$, in
which the hyperboloid turns into a paraboloid with the radius of curvature
equal to $R$.  In this limit we obtain from
eq.~(\ref{eq:opticalA}), $A_{\rm OPT}(0)=1$, whereas the PFA gives
$A_{\rm PFA}(0)=0$, so the predictions for the paraboloid (hence
the superscript \emph{para}) differ dramatically,
\begin{eqnarray}
f^{para}_{\rm OPT}(h/R,0)&=&1+h/R+ {\cal O}(h^2/R^2),\\
f^{para}_{\rm PFA}(h/R,0)&=&1 + {\cal O}(h^2/R^2) \, .
\end{eqnarray}

Experimenters measure forces, not energies.  The predictions for the
Casimir
force can be read off the figures for each of the approximations.
The correspondence is
particularly simple when $f(h/R,\theta)$ is approximated
to linear order in $h/R$, as in eq.~(\ref{eq:defA}),
\begin{equation}
{\cF}=-\frac{d\cE}{dh}=-\frac{\pi^3R}{720h^3}\left(
1+\frac{h}{2R}A(\theta)+...\right) \, .
\label{eq:force}
\end{equation}
Another limit that is interesting in principle but cannot be
analyzed within our approximation is the `cusp' limit in which
$R\to 0$ while $\theta$ is held fixed. Here, however, we face a
major difficulty, since $h$ measured in units of $R$ is going to
infinity, corresponding to the far right in Fig.\ \ref{fig:f}.
Since $h$ is the measure of the wavelengths that dominate the mode
sum in eq.\ (\ref{eq:casimiren}), the cusp limit is dominated by
long wavelengths and diffraction (which is ignored in the optical
approximation to the propagator) becomes more and more important.

\subsection{Finite Plate Studies}
\label{sec:finite}

The optical approximation allows one to study the effect of finite
bounding surfaces.  In the case of a hyperboloid it is not hard to
extend our algorithm to the case when the plate is replaced by a
finite disk of radius $L$.  Since this is a configuration that may
well be possible to examine experimentally, we work out the
predictions of the optical approach and compare them with the PFA
and the semi-classical approximation.  In order to simplify the
analysis we fixed $\theta=30^\circ$, though, of course, any other
value of $\theta$ can be analyzed as well.

It is necessary to restrict ourselves to situations where $h\ll
L,R$, in order to being able to neglect edge diffraction effects.
However no restriction is posed on the relative magnitude of $R$
and $L$.  In particular, the transition between $L>R$ and $R<L$
can be studied.  One can think of the optical approximation as a
semi-classical approximation to the Casimir energy for $h\ll L,R$
\emph{uniformly valid} as a function of the parameter $R/L$ while
the semi-classical approximation breaks down when $R/L\gtrsim 1$.
This issue has been discussed in more general terms in paragraph
III.C of Ref.~\cite{opt1}.

If we factor out the most divergent term we can write
\begin{eqnarray}
\cE=-\frac{\pi^3 R}{1440 h^2}g(h/R,L/R,\theta),\nonumber
\end{eqnarray}
which is related to the function $f$, previously defined, by
\beq
f(h/R,\theta)=\lim_{L/R\to\infty}g(h/R,L/R,\theta) \, .
\eeq
It is clear on physical grounds that if the radius of the finite
plate, $L$, is small compared to the radius of curvature of the
hyperboloid then the curvature of the hyperboloid can be
neglected.  In this case the result must reduce to that obtained
for two parallel plates, one infinite and one of radius $L$,
\beq
\cE=-\frac{\pi^{3}L^{2}}{1440h^{3}} \, .
\eeq
 This implies that for $L/R\ll 1$ asymptotically one must find
\begin{equation}
    g(h/R,L/R,\theta)\sim
    \frac{L^{2}}{hR} \, .
        \label{smallplate}
\end{equation}

To gain a more complete understanding of the different regimes for
varying $L/R$ and  $h/R$ a good starting point is again the PFA,
which has, as it will turn out,
  the same qualitative behavior as the optical approximation though it
  differs quantitatively. According to the PFA, for finite plate we
  have (here $\eta\equiv h/R,\ \lambda\equiv L/R$)
\beq
    g_{\rm PFA}(\eta,\lambda,\theta)=\frac{\eta \,{\lambda
        }^2 + {\cot^2 \theta }\, \left( {\lambda }^2 + 2\,\cot \theta
        \, \left( \cot \theta - {\sqrt{{\Mfunction{\lambda }}^2 +
        {\cot^2 \theta }}} \right) \right) }{{\left( \eta + {\cot^2
        \theta }\,\left( -1 + {\sqrt{1 + {\lambda }^2\,{\tan^2 \theta
        }}} \right) \right) } ^2} \, .
  \label{unexpanded}
\eeq
The expansion of this function in powers of $\eta$,
\begin{eqnarray}
   g_{\rm PFA}(\eta,\lambda,\theta) &=& 1 + \eta\,{\tan^{2} \theta
    }\,\nonumber\\ &-& \eta^{2} 
\frac{\left( 6\,\cot^2 \theta - 
6\,\cot \theta \,{\sqrt{\cot^2 \theta + {\lambda }^2}} + 
{\lambda }^2\,\left( 5 - 
2\,{\sqrt{1 + {\lambda }^2 \tan^2 \theta }} \right) 
\right) }{\cot^6 \theta \,{\left( 1 - 
{\sqrt{1 + \lambda^2 \tan^2 \theta }} \right) }^4} + {\Mfunction{O}(\eta )}^3
  \label{expansion}
\end{eqnarray}
sheds considerable light on its non-uniform behavior.  For any
$\lambda$ (and $\theta$) it is possible to choose $\eta$ so small
that the situation reverts to the hyperboloid opposite an infinite
plate ({\it i.e.\/} the term of $\Mfunction{O}(\eta^{2})$ is
negligible) and eq.~(\ref{expansion}) reduces to
eq.~(\ref{eq:fpa}).  The fact that the coefficient of $\eta^{2}$
is proportional to $\lambda^{-4}$  when $\lambda \to 0$ signals
non-uniform behavior in $\eta$ and $\lambda$.  For small $\lambda$ the
domain of linear 
growth with $\eta$ continues only up to $\eta\sim\lambda^{2}$
where $g$ has a maximum.  For larger $\eta$ the expansion of
eq.~(\ref{unexpanded}) in powers of $\eta$  breaks down and the
small $\lambda$, finite parallel plate limit of
eq.~(\ref{smallplate}) applies, so expanding $g$ in powers of
$\lambda^2/\eta$ we find
\beq
g_{\rm PFA}\sim\frac{{\lambda }^2}{\eta } - \frac{3\,{\lambda
}^4}{4\,{\eta }^2} +
  {\Ord{\frac{\lambda^6}{\eta^3}}},
\eeq
whose first term, once $h, R$ and $L$'s have been restored,
coincides with eq.~(\ref{smallplate}). For $\lambda\lesssim 1$ the
achievement of the maximum in $\eta$ represents the set-in of the
parallel plate limit:\ {\it i.e.\ }$R$ is large with respect to
$L$ and $h$ is large enough that the contribution to the energy is
not too much concentrated near the tip. When $\lambda \gtrsim 1$
even though we cannot neglect the curvature over distances of
order $L$ the contribution to the energy is spread enough so that
the parallel plates approximation works again. However in this
region we expect curvature effects (not captured by PFA) to be
non-negligible. Here hence we expect -- and we find -- the biggest
differences between the optical data and PFA.

The semi-classical approximation does not predict any change in the
energy with the plate radius $L$ and hence cannot predict the
parallel plates limit.  This is due to the fact that the only
semi-classical contribution comes from the periodic orbit bouncing
back and forth from the tip of the hyperboloid to the plate and
this ignores completely the transverse radial direction.  This
becomes pathological in the case when the geometry reduces to that
of two parallel plates, where it gives completely wrong results.
Explicitly, it predicts an energy $\cE\propto\hbar c R/h^2$ while
the correct result is independent of $R$.\footnote{One can see
this as the result of inverting two limits.  The parallel plates
case is obtained by taking $R\to\infty$ \emph{before} $h\to 0$
while the semi-classical approximation takes $h\to 0$
\emph{before} $R\to\infty$.}

\begin{figure}
\includegraphics[width=12cm]{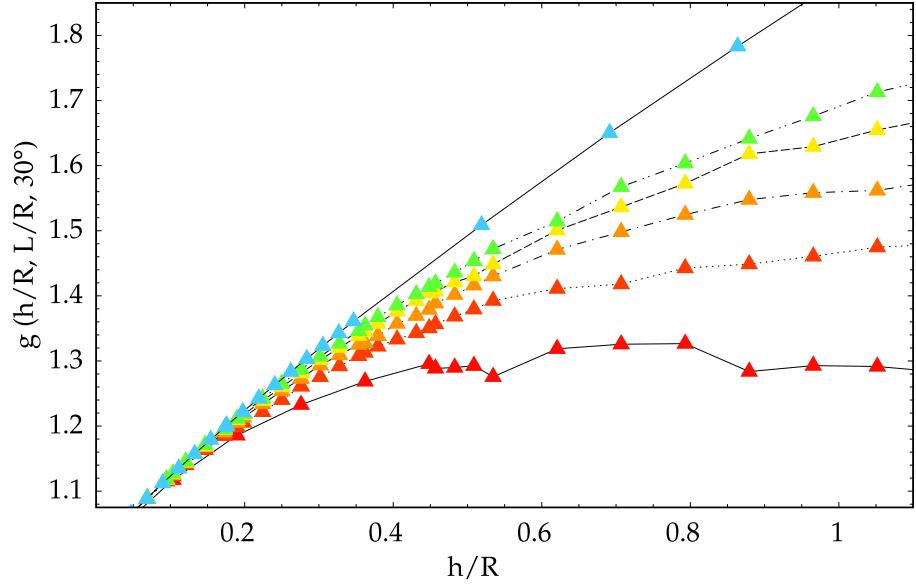}
\caption{\label{fig:finiteplate} \sl (Color online) The function
$g(h/R,L/R,\theta)$ given by the optical approximation. $\theta =
30^\circ$ for all of the curves. From down to up (red to green)
the first five set of points have $L/R=$ $1.5\sqrt{3}$,
$2.0\sqrt{3}$, ..., $3.5\sqrt{3}$,  the blue triangles
(uppermost curve) are the
infinite plate case. The wiggles for large $h/R$ and small $L/R$
are related to the accuracy of our computation, and are smoothed
out to a large extent when we increase the accuracy.}
\end{figure}

\begin{figure}
\includegraphics[width=10cm]{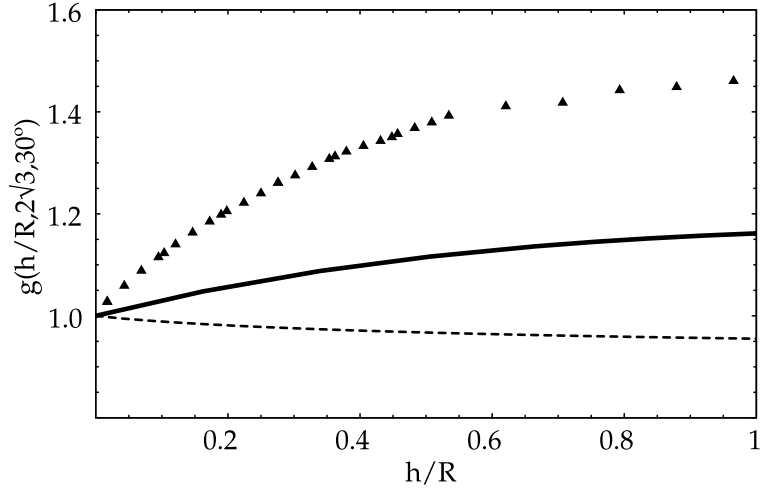}
\caption{\label{fig:ffpfa} \sl Comparison of $g(h/R,L/R,\theta)$
given by the optical approximation (triangles), the PFA
(continuous) and semi-classical result (dashed curve).
$\theta=30^\circ$ and $L=2\sqrt{3}R$ for all the curves.}
\end{figure}

With these considerations in mind we now turn to the results of
the optical approximation.  The optical approximation to $g$ is
shown for various values of $L/R$ as a function of $h/R$ in
Fig.~\ref{fig:finiteplate}.  The linear term in an expansion in
$h/R$ {at fixed $L/R$} (previously called $A$) does not depend on
$L$. This result is shared with the PFA (see eq.~(\ref{expansion}))
and the semi-classical approximation, which is completely independent
of $L$.  However, the general dependence on $h/R$ is completely
different.  This can be seen graphically in Fig.~\ref{fig:ffpfa}.

\begin{figure}
\includegraphics[width=11cm]{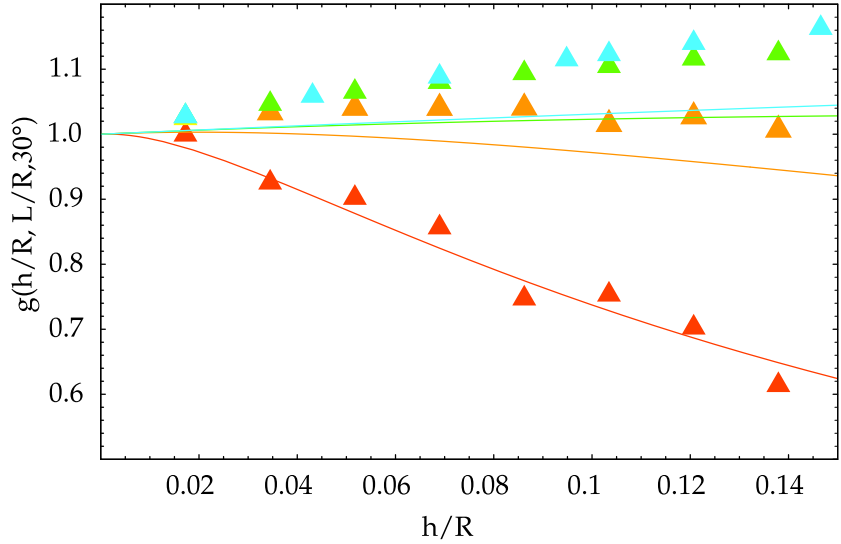}
\caption{\label{fig:ffinitePP} \sl (Color online) Comparison of the optical
approximation  and PFA for $g(h/R,L/R,\theta)$ for
$\theta=30^{\circ}$ and various $L/R$ as a function of $h/R$. From
down up (red to blue)
$L/R=\sqrt{3}/4,\sqrt{3}/2,\sqrt{3},2\sqrt{3}.$ The continuous
lines are the PFA predictions. The $L/R=\sqrt{3}/4$, red data are
fitted by the PFA within error bars (not shown) and the larger the
plate the worse the agreement.}
\end{figure}

Returning to Fig.~\ref{fig:finiteplate}, we see that the optical
approximation predicts variation of $g$ with $L/R$ for
$h/R\lesssim 1$.  The $L$ dependence of the PFA agrees
quantitatively with the optical approach for $h/R\ll 1$ and in the
parallel plate limit, where they both predict $g\to L^{2}/hR$.
However they differ in the intermediate range of $h/R$ and $L/R$.
In Fig.\ \ref{fig:ffinitePP} we compare the $L/R$ dependence
predicted by the optical approximation with the PFA. For the
smallest value  of $L/R$ ($L=\sqrt{3}R/4$), the two agree within
the error bars on the optical data both at very small $h/R$ and
larger $h/R$ where they both approach the ``parallel plate''
regime. This is the more evident manifestation of the uniform
validity of the optical approximation as the geometry is changed.
Notice that the agreement becomes worse as $L$ is increased. This
makes it clear that PFA only captures correctly a small region
around the tip of the hyperboloid, where the paths are almost
periodic. Even here it ignores the enlargement factor. Continuing
to increase $L$ at fixed $h$, we enter the region $h\ll L^2/R$,
where $g$ reduces to the infinite plate prediction and the slopes
of optical and PFA curves will differ according to the previous
discussion.

{The optical approximation, like the PFA, predicts a maximum
in $g(h/R,L/R,\theta)$ as a function of $h/R$ at fixed  $L/R$.}
The maximum of $g$ for fixed $L/R$ and $\theta$ is already evident
in Fig.~\ref{fig:finiteplate} for $L=1.5 \sqrt{3}R$.  {It
occurs} for every finite $L/R$ --- even though this cannot be seen
in Fig.\ \ref{fig:finiteplate}.  We will call this value
{$g^*(L/R, \theta)$, and the value of $h/R$ at which this is
found will be $\eta^*(L/R,\theta)$.}  For $L/R\gtrsim 1.5\sqrt{3}$
the maximum occurs in a region of $h/R$ that is beyond the
applicability of our approximation ($h^{*}/R\gtrsim 1$).

\begin{figure}
\includegraphics[width=11cm]{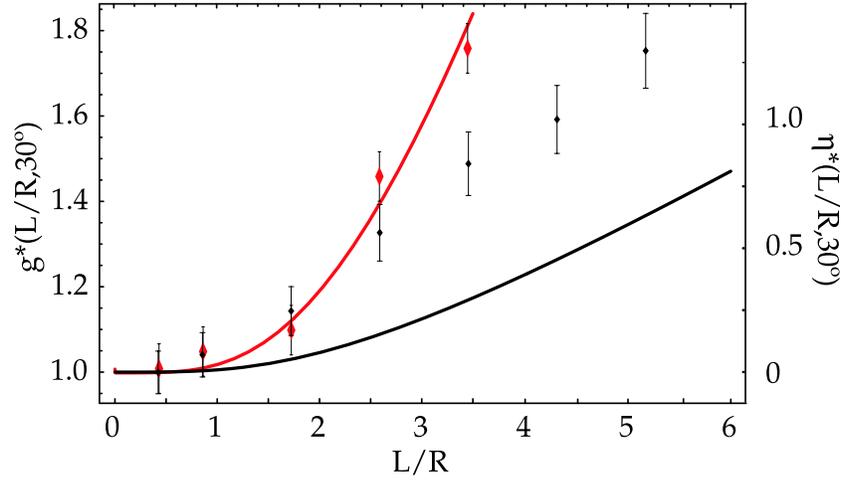}
\caption{\label{fig:fstarfinite} \sl {(Color online) The lower, black
    curve is the PFA prediction for $g^{*}(L/R,\theta)$, the maximum of
$g(h/R,L/R,\theta)$ as a function of $L/R$ and the black points
are the optical results. The upper, red curve is PFA prediction
for $h^{*}/R=\eta^*(L/R,\theta)$, the value of $h/R$ at which the
maximum occurs, as a function of $L/R$, and the red points are
optical data. Both curves have $\theta=30^\circ$. The optical
approximation data carry an error bar of $~5\%$ on $g^*$ and $0.1$
on $\eta^*$, estimated from the accuracy of the numerical
integration and the maximum finding procedure.}}
\end{figure}

We present the data for $g^*$ {and $\eta^{*}$} in Fig.\
\ref{fig:fstarfinite}.  The data on $\eta^{*}$ are in good
agreement with the PFA prediction.  The data on $g^*$, however,
are in worse accord. One could then say that PFA and the optical
approximation disagrees on the predictions for the energy (and
hence the force) but they agree in identifying the basic length
scales of the problem.

In conclusion, there are also important differences among the
various approximations when applied to the finite plate case.
These differences are more marked than in the infinite plate case.
In particular the semi-classical approach \cite{SandS} does not
depend on the size of the plate (nor on that of the hyperboloid)
at all.  For finite $L$ the optical approximation data for $g$
reach a maximum and then decrease. This is not captured by the
semi-classical approximation and is understood by the PFA only
qualitatively but not quantitatively.

\section{Conclusions}

Studies of the geometrical dependence of Casimir forces are in their
infancy.  Experiments are just reaching the level of accuracy where
deviations from the naive proximity force approximation can be
detected.  There are few theoretical calculations for geometries other
than parallel plates.  The optical approximation offers hope for an
accurate estimate of Casimir forces for a wide range of geometries.
However we do not know how to bound the corrections to this
approximation.  The configuration of a hyperboloid and plate offers a
flexible laboratory for studying approximations.  It is likely to be
accessible to experiment.  One should keep in mind, however, that
actual experiments involve electromagnetic fields not scalar fields.
In the case of parallel plate the only modification is an increase of
the force by a factor of two.  For curved surfaces the effect is not
so well understood, but the dominant effect is still simply a factor
of two.

The goal of this paper has been to work out the predictions of the
optical approach so they can be compared with experiment and
contrasted with other approximations.  We find that the optical
approximation differs significantly from the PFA and the
semi-classical approximation.  The difference becomes more
important as the opening angle of the hyperboloid, $\theta$,
decreases.  All approximations agree on the first term in an
expansion in $h/R$, but differ thereafter.  We certainly expect
the optical approach to be more accurate, but in the absence of an
estimate of errors, only comparison with experiment or with a
numerical computation in the spirit of Ref.~\cite{Gies03} can
settle the issue.  We have also studied the effects of replacing
the infinite planar plate with a finite disk.  We found notable
differences both with PFA and the semi-classical approximation of
\cite{SandS}.  In some domains of the parameters $h,L,R$ these
differences are so relevant that we believe they can be easily
measured in an actual experiment.

More generally speaking, the high precision experiments to be
performed in the near future will be able to measure the
next-to leading order terms in a small distance expansion and will
hence be able to tell us whether the recent developments in the
theoretical analysis of Casimir effects for curved geometries
point in the right direction.

\section{Acknowledgements}

We thank G.~L.~Klimchitskaya and V.~M.~Mostepanenko for helpful
discussions. O.~S. gratefully acknowledges conversations with K.
Fukushima. O.~S. was supported by the \textit{Deutsche
Forschungsgemeinschaft} under grant DFG Schr 749/1-1. A.~S. is
Bruno Rossi Fellow and Jonathan A. Whitney Fellow, partially
supported by INFN.  This work is also supported in part by funds
provided by the U.S.\ Department of Energy (D.O.E.) under
cooperative research agreement DF-FC02-94ER40818.

\section{Appendix}

We have developed a C-program that allows one to calculate the optical
contribution to the Casimir energy.  In this Appendix we discuss the
algorithm in some detail because we believe it may be relevant for
other problems, such as the study of density of states oscillations in
chaotic billiards.

\subsection{Outline}

The starting point of the numerical computations in this paper is eq.\
(\ref{eq2}) which we repeat here for convenience:
\begin{equation}
        {\cal E}_{\rm optical}=-\frac{\hbar c}{2\pi^{2}}\sum_{\bf
        n}(-1)^{n} M_{n}\int_{{\cal D}_{\bf
        n}}d^{3}x\frac{\sqrt{\Delta_{\bf n}(x)}}{\ell_{\bf
        n}^{3}(x)} \, .
\end{equation}
From this equation, it is obvious which kind of question a
numerical program has to address. Performing the sum over ${\bf
n}$ is a trivial task since we consider only paths with a fixed
upper bound on the number of reflections, in our case six. The
second ingredient is a routine that performs the spatial
integration. Since the surfaces we have are cylindrically
symmetric, we introduce cylindrical coordinates, and carry out the
integration over $\varphi$. Then we are left with an integral over
$z$ and over $\rho$.  Both integrations are done using an adaptive
step size differential equation solver, in our case a slightly
modified version of {\tt odeint} \cite{NumericalRecipes}. The
integration routine will choose a number of points where the path
length $\ell_{\bf n}$ and the enlargement factor $\Delta_{\bf n}$
are required. In order to compute these quantities, the optical
paths are needed. These paths are closed paths of minimum length,
specified by the bulk point they start from, the number of
reflections and the sequence of surfaces they reflect from. The
requirement that they be of minimum length is equivalent to saying
that they are - between reflections - straight, and the
reflections are specular. The way we determine these paths will be
treated extensively below. Once an optical path is determined, it
is obviously trivial to determine its length. It is also a simple
matter to determine the enlargement factor, as will also be
described in some detail below. As a last point, it should be
mentioned that the determination of the integration domain is
rather implicit. Formally, we integrate over the volume enclosed
between the two surfaces (except for the one-reflection term
\cite{pap1}). The reduction of this volume to  ${\cal D}_{\bf n}$
comes about because for some points in the volume no closed path with
specular reflections only exists. Hence, if our routines for finding
minimum paths do not find any, the contribution of this point to the
integral is set to zero. 

\subsection{Minimum Paths and Subtleties}

\begin{figure}
\centerline{
\includegraphics[width=7cm,height=6cm]{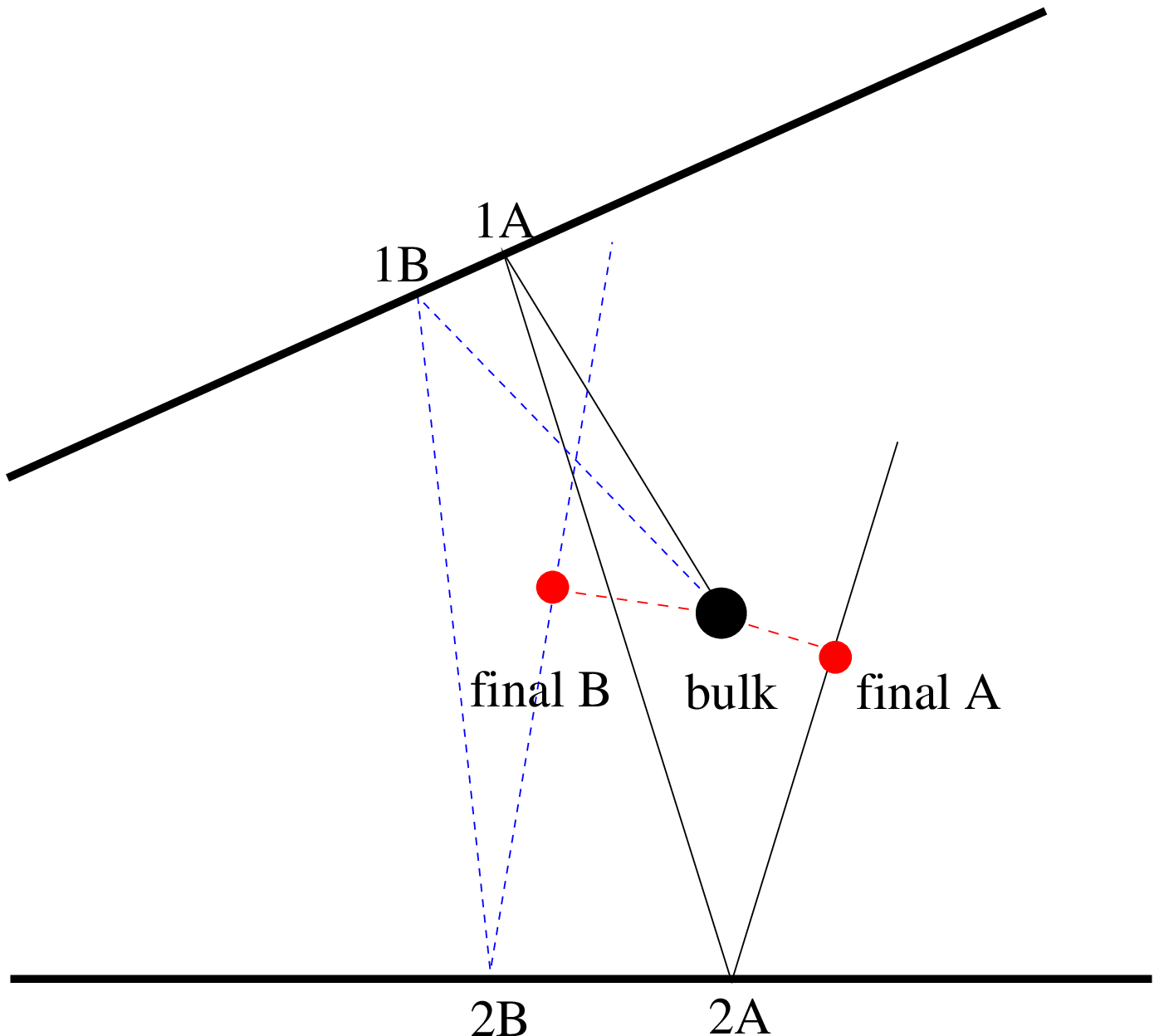}
\hskip1cm
\includegraphics[width=7cm,height=6cm]{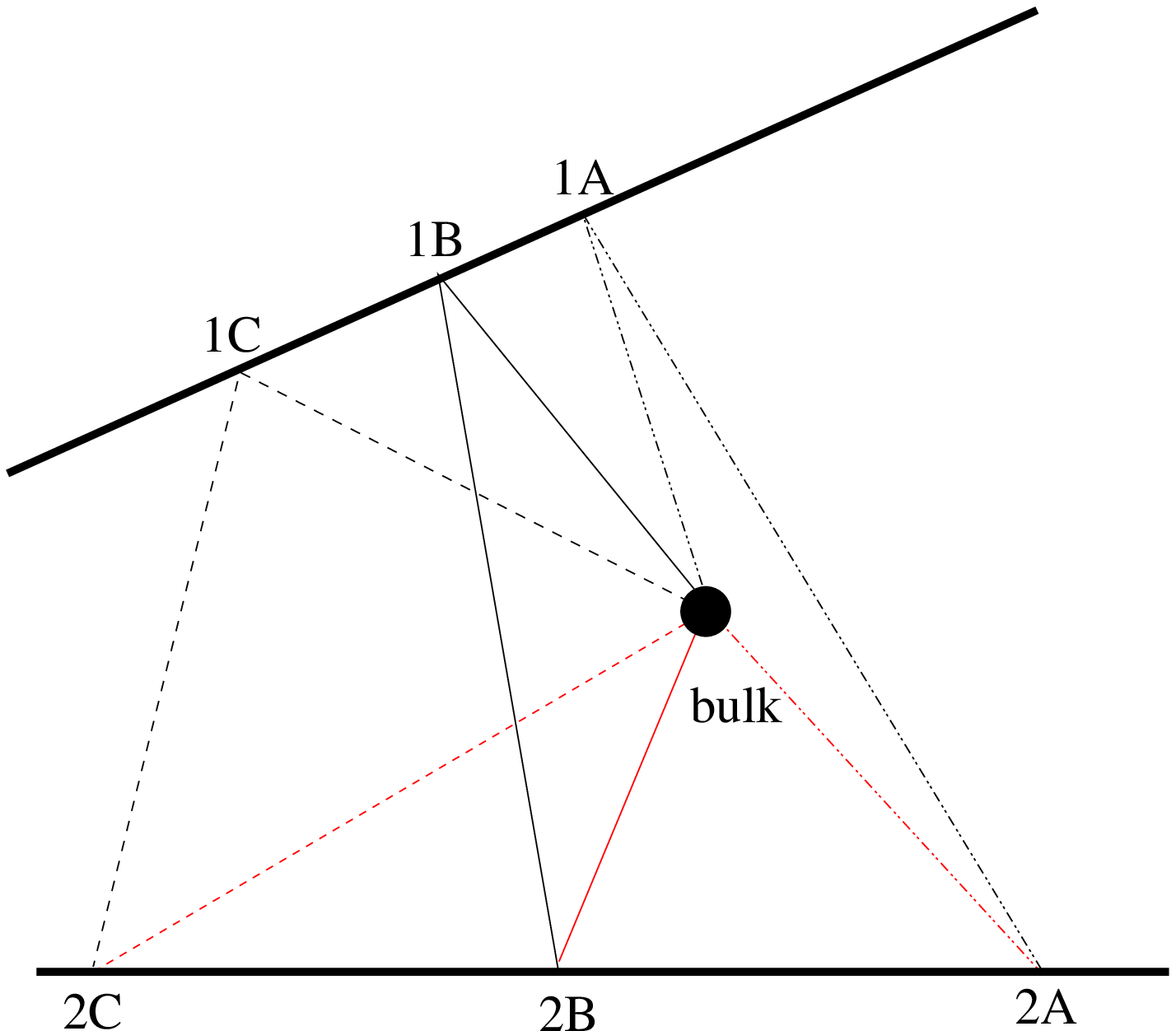}
  } \caption{\label{fig:min1}\sl (Color online) The left panel shows
  how the ``open  path approach'' works, the right panel shows the
  procedure in the ``path length minimization''.  In the left panel, all
  reflections have equal incoming and outgoing angles; in the right
  panel only the reflections at points 1A,1B,1C have equal incoming
  and outgoing angles.}
\end{figure}

The integration routine chooses points in the bulk where path length
and enlargement factor are required.  Since it is known that a path of
minimum length will have --- at each reflection --- incoming and
outgoing angles identical, the determination of a minimum length path
is a one parameter minimization problem.
There are (at least) two different approaches to determine the path of
minimum length, both of which are used in our numerical procedure.
Either, the $n$-reflection path under consideration is allowed to be
open, but all reflections are specular (``open path approach''), or
the path is required to be closed, but then the last reflection is not
required to be specular (``path length minimization approach''):

\begin{itemize}
  \item {\bf Open path approach} (cf.  left-hand panel of fig.
  \ref{fig:min1}): Here we consider the path generated by a
  sequence of specular reflections.  Such a path will, in general, not
  return to the point at which it originated. Consider the piece of
  path that is obtained after n reflections.  Then consider
  the point on this piece of path that has the minimum distance
  (labeled \textit{final A} and \textit{final B} in the figure) from
  the point in the bulk
  where this path originated from (labeled \textit{bulk} in the
  figure).  Minimize this minimum distance by varying the point of the
  first reflection (choices labeled 1A, 1B in the figure).
  The minimum of this function is zero, and if a zero is found the
  path is indeed a closed path where all reflections are specular.
  Numerically, it is easier to find a \textit{zero crossing} than a
  minimum, therefore it is useful to define a signed distance, i.e. a
  distance that has a notion of whether the last piece of the path
  passes above the bulk point or below; some details on this will be
  given below.  We use the abbreviation \textit{SDZC} (`signed
  distance zero crossing') for this method.

  \item {\bf Path length minimization} (cf.  right-hand panel of fig.
  \ref{fig:min1}): In this case we insist that
  the paths be closed but give up the constraint that all
  reflections have to be specular.  In general, if we insist on the
  path being closed all reflections can be chosen to be specular save
  the last one.  Thus we reflect the path $n-1$ times, and note where the
  $n^{\rm th}$ reflection would have occurred.  Then we minimize the
  length of the path as a function of the initial reflection point.
  We use the abbreviation \textit{PLM} (`path length minimization')
  for this method.
\end{itemize}

\begin{figure}
\centerline{
\includegraphics[width=5cm,height=5cm]{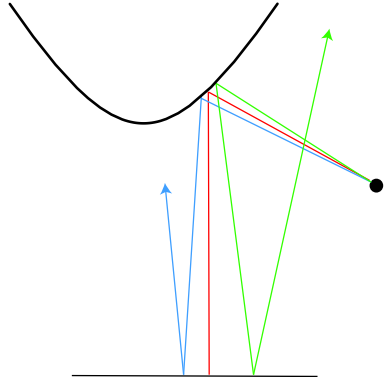}
  } \caption{\label{fig:ci1}\sl (Color online) This figure illustrates
  the subtlety described in the text regarding the possibility of a
  sign change of 
  the signed distance function without finding a closed path. The
  black dot indicates the bulk point. The middle path (red) is the
  path that detects the singularity in the sign function used.
}
\end{figure}

Both methods have advantages and disadvantages. The advantage of
the SDZC method is that if it works, then it works much faster,
since the zero crossings of the signed distance function are much
steeper than the minima of the path length function.  However, if
one wants to keep the determination of the sign in the SDZC method
simple, there are cases where -- as the initial point is varied --
the sign changes, but apparently not continuously, i.e.\ the
signed distance can not be made arbitrarily small. This
counter-intuitive behavior is best understood with a specific
example (see Figure \ref{fig:ci1}). In order to determine the sign
we follow the last part of the path till it has the same value of
$\rho$ as the bulk point where the path started from. Then we
compare the $z$ coordinate of this point on the path with the $z$
coordinate of the bulk point and we can say whether the path
passes above or below the bulk point. In order to understand how
one gets a sign change without finding a closed path, consider a
two-reflection path with the first reflection off the hyperbola
(situation in Figure \ref{fig:ci1}). If this path reflects off the
plate perpendicularly, it has found the singularity in our sign
prescription. Since this last part of the path can never have the
$\rho$ value of the bulk point, it cannot be decided whether it
passes above or below. This is the underlying cause for the sign
change without a zero crossing of the signed distance function: if
instead of this singular path we consider a path with initial
point on the hyperbola to the left (resp.\ right), the part of the
path reflecting off the plate will be reflected also to the left
(right) and hence passing \textit{below} (\textit{above}) the bulk
point.

If no closed path can be found using the \textit{SDZC} method we
use the \textit{PLM} instead. Both methods also have to deal with
difficulties that are not apparent in Fig.\ \ref{fig:min1}. There
we have shown two reflection paths only, and the problems appear
significantly first for four reflection paths.  First, a problem
for the \textit{SDZC} method is that given the initial point, an
$n$ reflection path might not exist if anywhere in between a piece
of path does not ``hit'' the designated next surface but simply
runs off to infinity.  Second, a problem for the PLM  is that
either the first section of the path or the last section of the
path may intersect one of the surfaces, thus rendering the path
illegal.  A useful method of handling these paths is to ensure
that they have a length that is (orders of magnitude) larger than
the largest ``correct'' length that can appear in the problem,
though still finite.  In the first case, this is ensured by
terminating the section of the path that does not hit a surface at
a very large distance, in the second case a large number is added
to the otherwise ordinarily computed path length.  The subtlety is
that on the one hand this large number should be much larger than
any path length occurring, so that an illegal path can be spotted
by simply looking at its length.  However, the number should not
be so big that -- with the prescribed numerical accuracy -- the
finite length information of the path length is lost.
The reason for this requirement is that for sufficiently many
reflections (and this is a serious problem already at six
reflections) almost all paths run off to infinity.
If the finite information is destroyed by the value we choose for
the large number, the \textit{PLM}  has no variation of path
length to work on. However, such a variation of the path length is
needed since the \textit{PLM} works in the following way: first,
path lengths for a finite number of initial points with $\rho$
values slightly above the $\rho$ value of the bulk point down to
$\rho=0$ are computed. Then within these points one searches for
the region where there has to be a minimum. In other word one
looks for three points $\rho_1< \rho_2< \rho_3$ with
$length(\rho_2) < length(\rho_3)$ and $length(\rho_2) <
length(\rho_1)$). Once this region is identified, the true minimum
is found by Golden Section Search \cite{NumericalRecipes}).
For example in {\tt double} precision C$++$, $10^{30}$ is too big,
whereas $10^{10}$ is just the right size for the large number.

\subsection{Enlargement Factors}

The computation of the enlargement factor is rather simple once
the first reflection point of a closed minimum path has been
found: take a step of unit length from the point in the bulk
towards the first point of reflection.  From the point thus
reached construct four new points: by going $\epsilon$ into
positive and negative $y$ direction (usually our computations take
place in the $x-z$ plane, therefore a step in $y$ and $-y$ is
guaranteed to be orthogonal), and a step each into the direction
orthogonal both to the $y$ axis and the direction where we took
our first unit step. These four points define four new paths: they
start at the original bulk point, and pass through these four new
points.  Then -- in case we are considering an $n$ reflection path
-- they are reflected $n$ times off the proper surfaces.  These
paths will not be closed since we consider only convex surfaces.
After $n$ reflections we determine the points of minimum distance
to the bulk point.  These four points together with the bulk point
determine four triangles. Their areas are added up to give $dA$,
whereas $d \Omega = 2 \epsilon^2$.

\begin{figure}
\centerline{
\includegraphics[width=8cm]{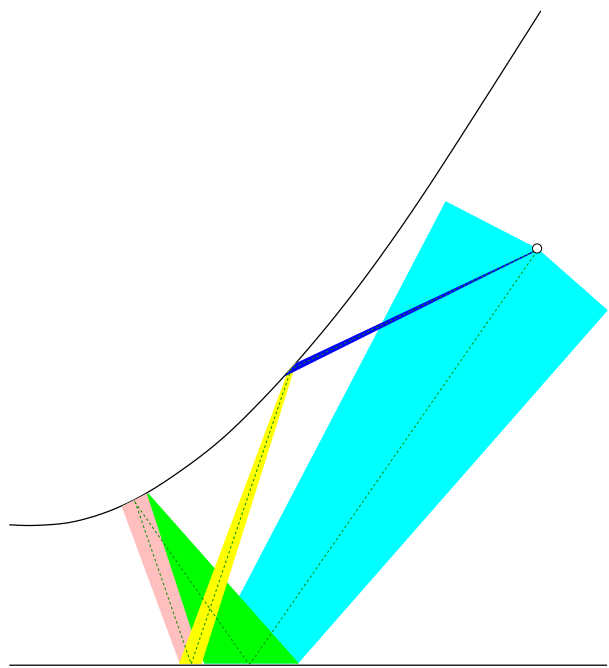}
}
\caption{\label{fig:enl}\sl (Color online) Computation of the
  enlargement factor for the example of a four reflection path. The
  dotted line indicates the closed minimum length path for which the
  enlargement factor is computed.}
\end{figure}

This procedure is numerically very convenient, since in order to
find the path of minimum length for each bulk point a couple of
hundred paths have to be computed, but once it is found, only four
more paths need to be computed for the enlargement factor (Fig. \ref{fig:enl}).

We have tested that this procedure produces stable results for
generic bulk points for widely varying values of $\epsilon$
between $10^{-8}$ and  $10^{-3}$.

\pagebreak

\end{document}